\documentclass[twocolumn,preprintnumbers,amsmath,amssymb,prl]{revtex4}

\usepackage[english]{babel}
\usepackage{graphicx}
\usepackage{bm}

\begin{document}

\title{Supercritical Gr\"{u}neisen parameter and its universality at the Frenkel line}

\author{L. Wang}
\affiliation{School of Physics and Astronomy Queen Mary University
of London, Mile End Road, London E1 4NS, United Kingdom}

\author{Yu. D. Fomin}
\affiliation{Institute for High Pressure Physics, Russian Academy
of Sciences, Troitsk 108840, Moscow, Russia \\ Moscow Institute of
Physics and Technology, Dolgoprudny, Moscow Region 141700, Russia}

\author{V. V. Brazhkin}
\affiliation{Institute for High Pressure Physics, Russian Academy
of Sciences, Troitsk 108840, Moscow, Russia}

\author{M. T. Dove}
\affiliation{School of Physics and Astronomy Queen Mary University
of London, Mile End Road, London E1 4NS, United Kingdom}

\author{K. Trachenko}
\affiliation{School of Physics and Astronomy Queen Mary University
of London, Mile End Road, London E1 4NS, United Kingdom}

\begin{abstract}
We study thermo-mechanical properties of matter at extreme conditions deep in the supercritical state, at temperatures exceeding the critical one up to four orders of magnitude. We calculate the Gr\"{u}neisen parameter $\gamma$ and find that it decreases with temperature from 3 to 1 on isochores depending on the density. Our results indicate that from the perspective of thermo-mechanical properties, the supercritical state is characterized by the wide range of $\gamma$ which includes the solid-like values - an interesting finding in view of the common perception of the supercritical state as being an intermediate state between gases and liquids. We rationalize this result by considering the relative weights of oscillatory and diffusive components of the supercritical system below the Frenkel line. We also find that $\gamma$ is nearly constant at the Frenkel line above the critical point and explain this universality in terms of pressure and temperature scaling of system properties along the lines where particle dynamics changes qualitatively.
\end{abstract}

\maketitle

\section{Introduction}

Dimensionless quantities play an important role in describing physical phenomena. One such parameter, the Gr\"{u}neisen parameter (GP), has been proved to be very useful in the theory of lattice vibrations and thermodynamics of solids. In solid state physics, the Gr\"{u}neisen parameter describes the change of system's elastic properties in response to volume change \cite{girifalco}:

\begin{equation}
\gamma=-\left(\frac{\partial\ln\omega}{\partial\ln V}\right)_T
\label{grun-def-w}
\end{equation}

\noindent where $\omega$ is the effective average frequency of particle vibrations, $V$ is the system volume.

The Gr\"{u}neisen parameter can also be related to system energy and pressure \cite{vocadlo}:

\begin{equation}
\gamma=V\left(\frac{\partial P}{\partial E}\right)_V
\label{grun-def}
\end{equation}

Eqs. (\ref{grun-def-w}) and (\ref{grun-def}) are equivalent in the condensed matter systems, but the second equation is more general and applies to gases, high-temperature fluids and plasma where individual particles do not vibrate. Eq. (\ref{grun-def}) leads to \cite{vocadlo}

\begin{equation}
\gamma= \frac{\alpha_P B_T V}{C_V}
\label{grun-def1}
\end{equation}

\noindent where $\alpha_P$ is the thermal expansion coefficient, $B_T$ is the isothermal bulk modulus and $C_V$ is the constant volume heat capacity.

As follows from (\ref{grun-def-w}) and (\ref{grun-def}), $\gamma$ is a thermo-mechanical quantity that is important for thermo-mechanical effects, in particular for those involving extreme temperatures and pressures. These include shock wave effects, rapid expansion and heating of systems absorbing nuclear radiation and so on. Here, the GP becomes particularly important: if, as if often the case, the pulse duration is shorter than the timescale of acoustic transport, the induced thermal pressure is directly proportional to $\gamma$. Consequently, the GP is extensively used in analyzing the equations of state of condensed matter and plasma at extreme conditions.


For most condensed matter systems, the range of $\gamma$ is $0.5-4$. Diamond is an  ``ideal" Gr\"{u}neisen system with $\gamma=1$ \cite{occelli}. Systems with large pressure derivatives of $B$ (lattice stiffens quickly with compression) often have large $\gamma$ \cite{vocadlo}. Interestingly, since $B_T$ and $C_V$ are positive in equilibrium, the sign of $\gamma$ is governed by the sign of $\alpha_P$. Some systems such as Cu$_2$O and ScF$_3$ have small negative $\gamma$ in a quite large temperature and pressure range \cite{Martin}, accompanied by negative $\alpha_P$ and softening of force constants on compression. Negative $\gamma$ can also be seen in shock-wave experiments due to non-equilibrium smeared phase transformations \cite{brazhkin-ufn}.

Compared to solids, relatively little is known about the GP in liquids and dense gas states. For the ideal gas, $\gamma=\frac{2}{3}$ is a constant as follows from $E=\frac{3}{2}PV$. The same result also applies to the degenerate electron gas \cite{Gosp}. For the frequently discussed hard-spheres model, $\gamma$ can be calculated from the Carnahan-Starling equation $Z=\frac{PV}{Nk_BT}=\frac{1+\eta+\eta^2-\eta^3}{(1-\eta)^3}$, where $\eta=\frac{\pi}{6}\rho\sigma ^3$ is the packing fraction of hard spheres of diameter $\sigma$ at density $\rho$ \cite{car-starling}. This gives $\gamma=\frac{2}{3}f(\rho)$, where $f(\rho)$ is a function of density, implying that the GP of hard spheres is constant along isochores. For model Van del Waals system, $\gamma=\frac{2}{3}\times\frac{V}{V-Nb}$, where $ b$ is the cohesion volume, the GP diverges when the volume becomes close to the critical volume \cite{Mariano}. The soft-sphere interaction with weak attraction modifies the GP, and there are analytical evaluations of this effect \cite{shaner1,shaner2}. Based on certain assumptions and in reasonable agreement with simulations of noble-gas systems \cite{lurie}, there are numerical evaluations of the GP for the commonly-used Lennard-Jones potential \cite{krivtsov}. For more complicated liquids such as water and mercury, the GP was calculated using (\ref{grun-def1}) and was found to increase with pressure, in contrast to its usual decrease in crystals \cite{knopoff}. The GP was also calculated in liquid Ar in a small range of pressure and temperature and was found to decrease on isobaric heating \cite{kor-tandon}. In a wider temperature and pressure range, $\gamma$ in Ar in the dense gas and liquid state increases on isothermal compression and is nearly constant on isochoric heating \cite{amoros}. $\gamma$ was also calculated from ensemble averages of fluctuations \cite{fluc}. Finally, $\gamma$ was evaluated using the radial distribution function of liquids with acceptable errors \cite{emampour}.

Notably, no studies or evaluations of $\gamma$ were done significantly above the critical point of matter. Supercritical fluids started to be widely deployed in many important industrial processes \cite{sup1,sup2} once their high dissolving and extracting properties were appreciated. Theoretically, little is known about the supercritical state, apart from the general assertion that supercritical fluids can be thought of as high-density gases or high-temperature fluids whose properties change smoothly with temperature or pressure and without qualitative changes of properties. This assertion followed from the known absence of a phase transition above the critical point. We have recently proposed that this picture should be modified, and that a new line, the Frenkel line (FL), exists above the critical point and separates two states with distinct properties \cite{ropp,pre,ufn,prl,phystoday}.


The main idea of the FL lies in considering how particle dynamics changes in response to pressure and temperature. Frenkel previously proposed that particle dynamics in the liquid can be separated into solid-like oscillatory and gas-like diffusive components and introduced liquid relaxation time $\tau$ as the average time between particle jumps between neighbouring quasi-equilibrium particle positions \cite{frenkel}. We proposed that this separation applies equally to supercritical fluids as it does to subcritical liquids: increasing temperature reduces $\tau$, and each particle spends less time oscillating and more time jumping; increasing pressure reverses this and results in the increase of time spent oscillating relative to jumping. Increasing temperature at constant pressure (or decreasing pressure at constant temperature) eventually results in the disappearance of the solid-like oscillatory motion of particles; all that remains is the diffusive gas-like motion. This disappearance represents the qualitative change in particle dynamics and gives the point on the FL. Notably, the FL exists at arbitrarily high pressure and temperature, as does the melting line. Quantitatively, the FL can be rigorously defined by pressure and temperature at which the minimum of the velocity autocorrelation function (VAF) disappears \cite{prl}. Above the line defined in such a way, velocities of a large number of particles stop changing their sign and particles lose the oscillatory component of motion. Above the line, VAF is monotonically decaying as in a gas \cite{prl}. Another criterion for the FL which is important for our discussion of thermodynamic properties and which coincides with the VAF criterion is $c_v=2k_{\rm B}$ \cite{prl}. Indeed, the loss of solid-like oscillatory component of motion implies the disappearance of solid-like transverse modes which, in turn, gives $c_v=2k_{\rm B}$ \cite{ropp}.

The qualitative change of particle dynamics and $c_v=2k_{\rm B}$ at the FL are two important insights that we will use below to discuss the universality of the GP at Frenkel line.

The aim of this paper is to calculate and analyze the GP deep in the supercritical state. We calculate $\gamma$ for two common model systems at temperature and pressure exceeding the critical ones by orders of magnitude. We find that $\gamma$ decreases with temperature from its solid-like to gas-like values on isochores. This implies that from the perspective of thermo-mechanical properties, the supercritical state is characterized by the range of $\gamma$ which includes the solid-like values. This is an interesting finding in view of the common perception of the supercritical state as being an intermediate state between gases and liquids, which we rationalize in terms of the relative weights of the oscillatory and diffusive components of particle motion. We also find that $\gamma$ is nearly constant at the Frenkel line in the supercritical state. We explain this universality in terms of pressure and temperature scaling of system properties along the lines where particle dynamics qualitatively changes.

\section{Simulation details}

Firstly, we use the molecular dynamics (MD) simulation package DL\_POLY \cite{dlpoly} to simulate the LJ  model. The simulated systems have $8000$ particles with periodic boundary conditions and the interatomic potential for Argon is the pair Lennard-Jones potential \cite{ar}. We have simulated 5 densities: $\rho=1.20$ g/cm$^3$, $1.35$ g/cm$^3$, $1.50$ g/cm$^3$, $1.90$ g/cm$^3$ and $2.20$ g/cm$^3$. The temperature in each simulation varies from melting temperature at the corresponding density up to $10000$ K with the interval 10 K. The MD systems were first equilibrated in NVE ensemble for 40 ps. The data were subsequently collected at different temperatures for each density and averaged over the period of 60 ps.

We have also simulated the soft-sphere system in a wide range of density and temperature. The soft-sphere interaction potential is $U(r)=\varepsilon \left( \frac{\sigma}{r} \right)^n$, where $n$ is the softness parameter. We have considered $n=6$ and $n=12$, respectively. For $n=6$, we performed MD simulations of energy and pressure and calculated $\gamma$ using Eq. (\ref{grun-def}). This part of the simulation work was performed using the LAMMPS MD package \cite{lammps}. A system of $4000$ particles in a cubic box with periodic boundary conditions is simulated. The reduced densities of the system are $\rho_1$*$=1.0$ and $\rho_2$*$=1.5$ and the temperatures vary from $T$*$=2.7$ to $3.4$ in the soft-sphere units. The Frenkel temperature of this system at this density is $T_{\rm F}$*$=3.1$. The equilibration and production runs involved $10^6$ steps with a timestep was set to $0.0001$.

\section{Results and Discussion}

\begin{figure}
\begin{center}
{\scalebox{0.7}{\includegraphics{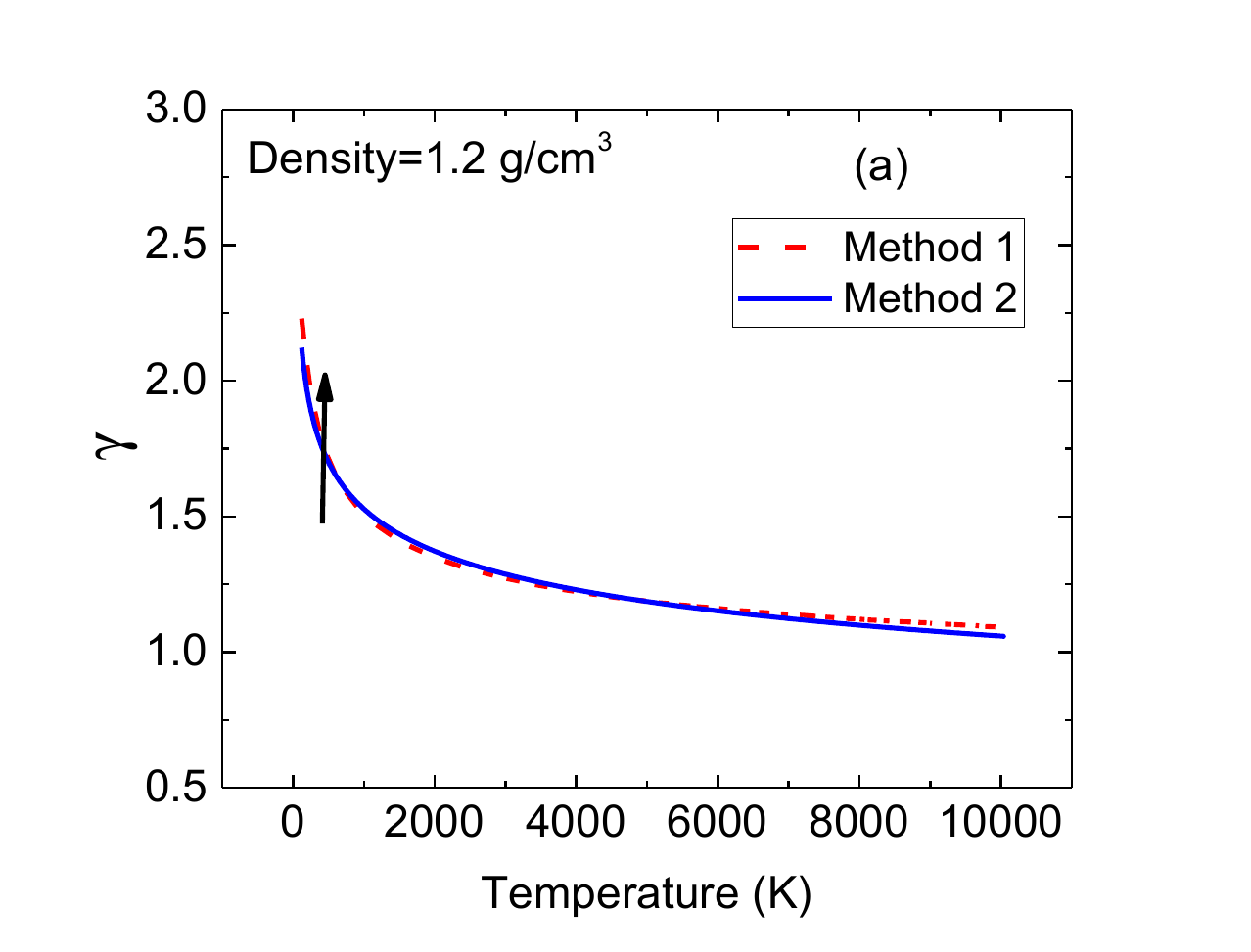}}}
{\scalebox{0.7}{\includegraphics{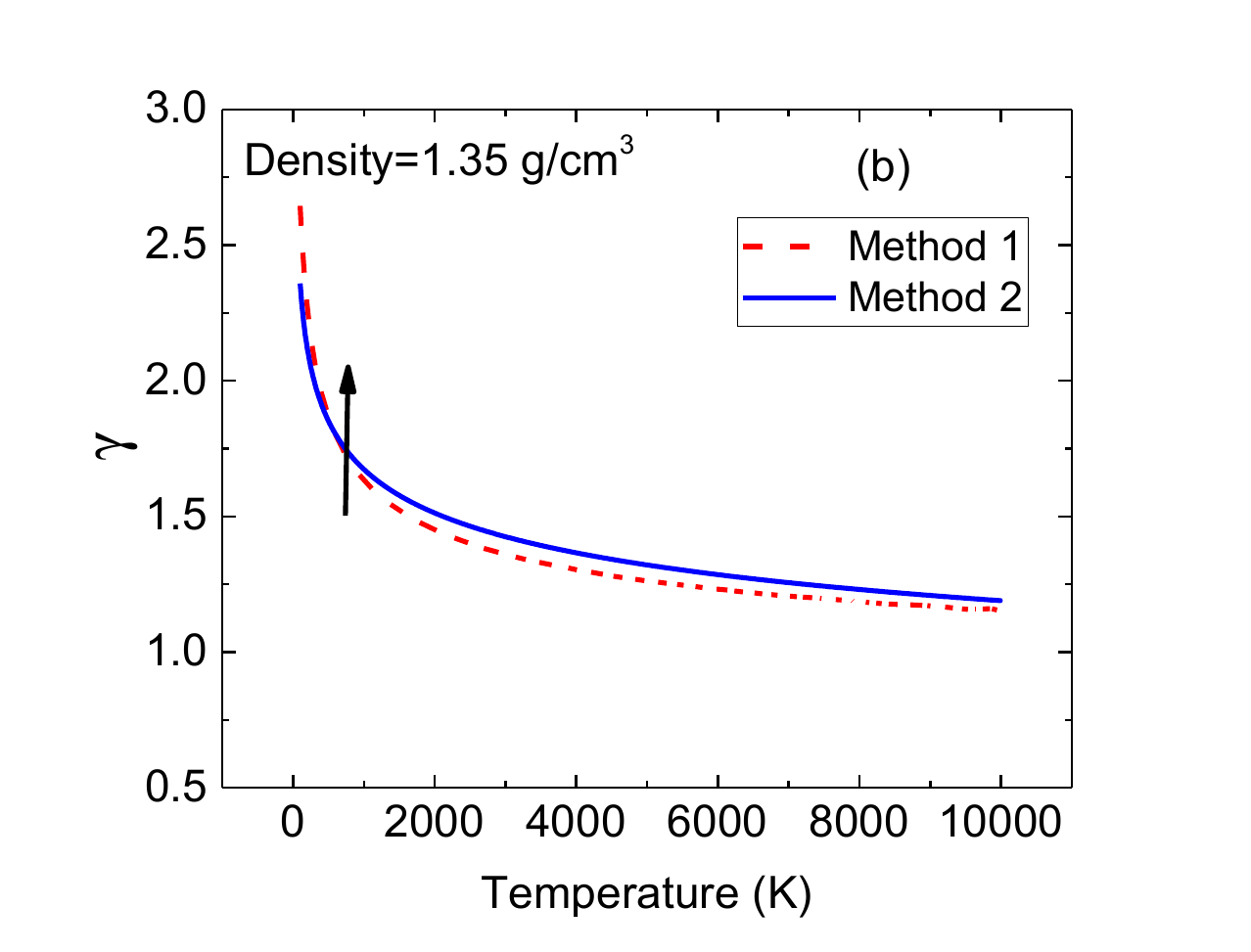}}}
\end{center}
\caption{Gr\"{u}neisen parameters calculated for the Lennard-Jones (Ar) system at two lower densities. The red dashed lines and blue solid lines are calculated using the two methods described in text. The arrows show the temperature at the Frenkel line.}
\label{low}
\end{figure}

\begin{figure}
\begin{center}
{\scalebox{0.7}{\includegraphics{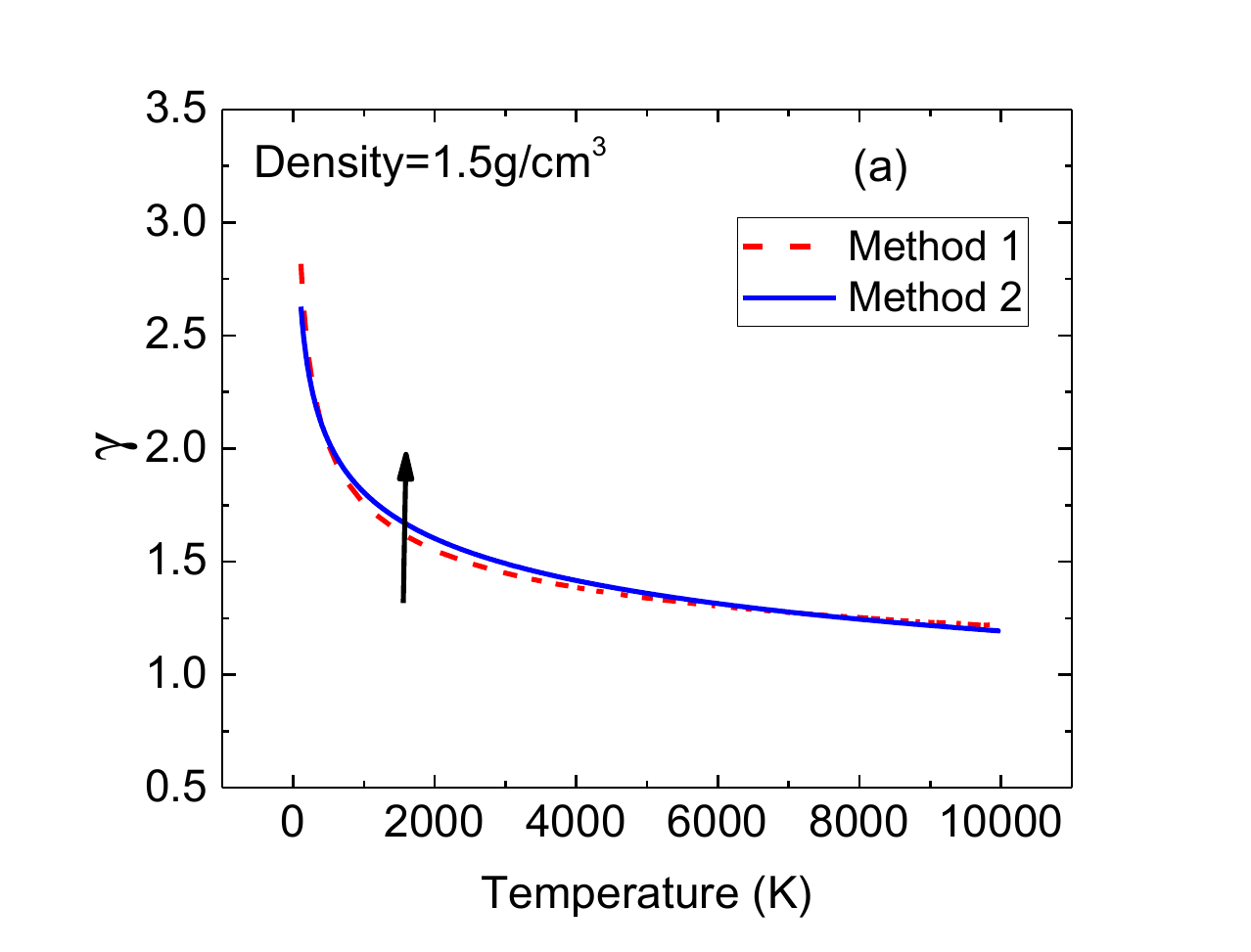}}}
{\scalebox{0.7}{\includegraphics{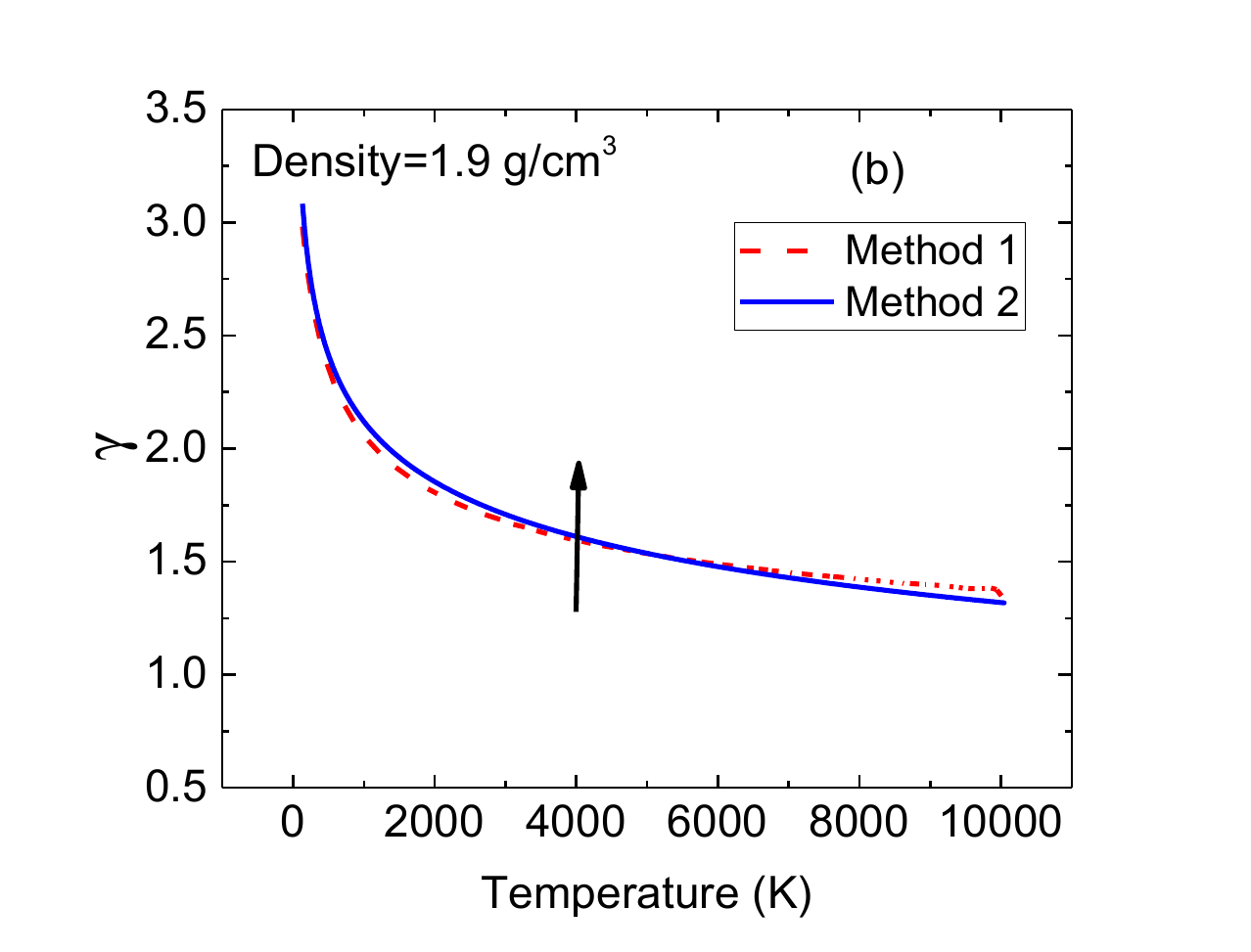}}}
{\scalebox{0.7}{\includegraphics{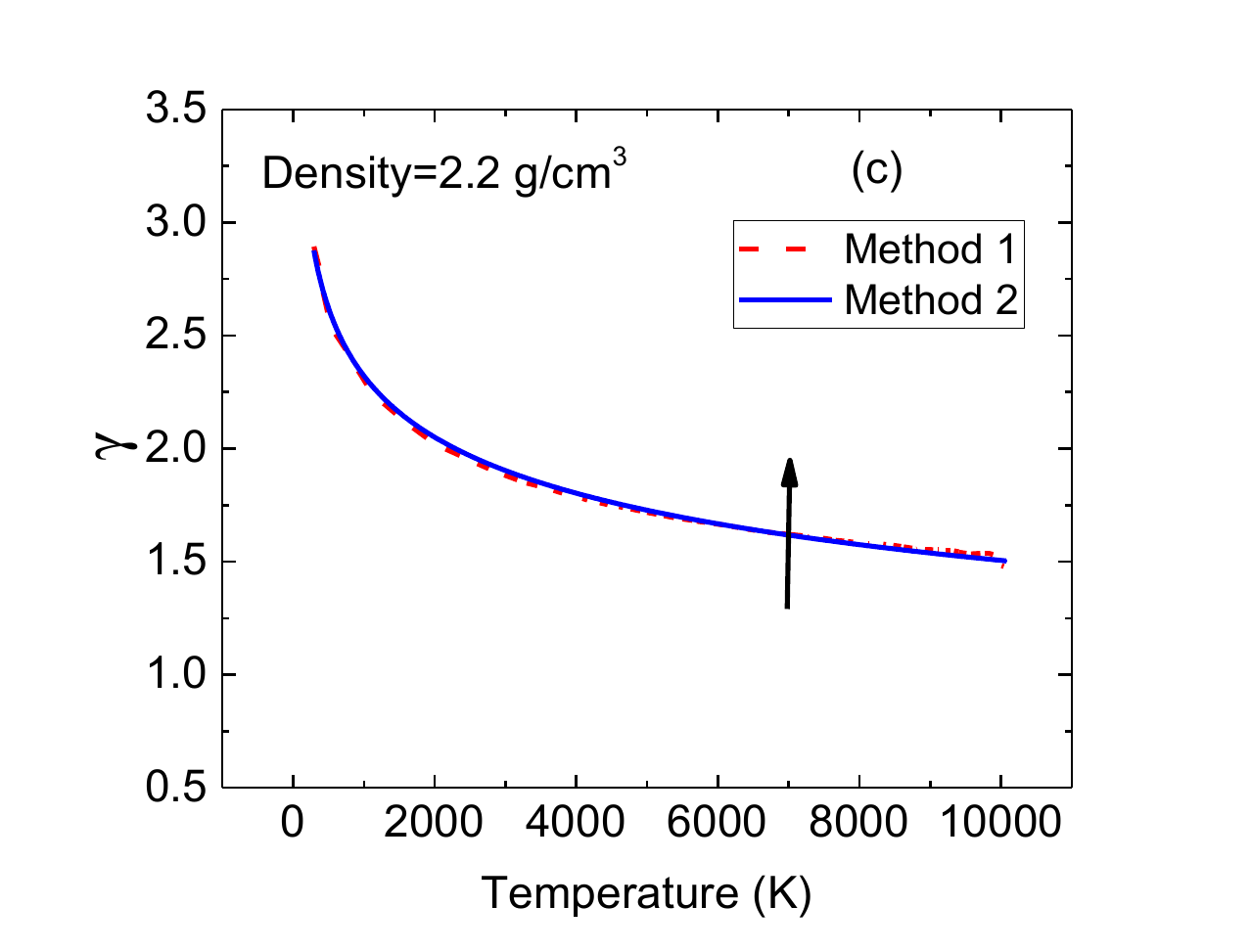}}}
\end{center}
\caption{Gr\"{u}neisen parameters calculated for the Lennard-Jones (Ar) system at three higher densities. The red dashed lines and blue solid lines are calculated using the two methods described in text. The arrows show the temperature at the Frenkel line.}
\label{high}
\end{figure}

We have calculated $\gamma$ using two methods. In the first method, we use $V$, $P$ and $E$ from the MD simulations, calculate $\gamma$ using Eq. (\ref{grun-def}) and fit the resulting values to the polynomial. In the second method, we first fit $V$, $P$ and $E$ to respective polynomials and then calculate $\gamma$ using Eq. (\ref{grun-def}). Both methods result in close curves for $\gamma$ as follows from Figures \ref{low} and \ref{high} discussed below.

We show the $\gamma$ calculated for Ar using both methods along 5 different isochors in Figure \ref{low} and Figure \ref{high}. We note that the range of thermodynamic parameters we used is record-high: the highest temperature and pressure exceed the critical ones by over one to two orders of magnitude. At each density, the arrow shows the corresponding temperature of the FL.

We observe that $\gamma$ calculated by Eq. (\ref{grun-def}) decreases from 2.5 to 1 with temperature at low density and from about 3 to 1.2 at high density. Notably, $\gamma=2.5-3.5$ are characteristic of the solid state. Therefore, our results indicate that from the perspective of thermo-mechanical properties, the supercritical state is characterized by the range of $\gamma$ which includes the solid-like values. This is an interesting finding in view of the common perception of the supercritical state as being an intermediate state between gases and liquids \cite{sup1,sup2}.

The solid-like values of supercritical $\gamma$ at low temperature can be explained by considering the relative weight of the oscillatory and diffusive components of motion in the supercritical state. This weight can be quantified by the $R$-parameter \cite{ropp}:

\begin{equation}
R=\frac{\omega_{\rm F}}{\omega_{\rm D}}
\end{equation}

\noindent where $\omega_{\rm F}=\frac{1}{\tau}$ and $\omega_{\rm D}$ is Debye frequency.

Recall that the oscillatory component of particle motion disappears at the Frenkel line. However, if the supercritical system is sufficiently below the Frenkel line, particles spend most of their time oscillating, and diffusive jumps between the quasi-equilibrium positions are rare. This gives $R\ll 1$. It is easy to show \cite{ropp} that in this case the average system energy is well approximated by the energy of the oscillatory motion. Therefore, basic thermodynamic properties of the supercritical system below the FL are solid-like, as are the dynamical properties related to phonons. Hence we expect $\gamma$ to be characterized by the solid-like values in this regime.

We can explore the similarity between $\gamma$ of the supercritical systems below the FL and their solid-like values further, by using the solid-like equation (\ref{grun-def-w}). We have earlier evaluated Debye frequencies $\omega_{\rm D}$ for the LJ system for two supercritical densities below the FL: $\omega_{\rm D}=7.2$ THz for $\rho=1.50$ g/cm$^3$ and $\omega_{\rm D}=18.4$ THz for $\rho=1.90$ g/cm$^3$ \cite{ling}. Using these values and $\omega\propto\rho^\gamma$, which follows from (\ref{grun-def-w}), gives $\gamma\approx 3.8$. This is in reasonable agreement with $\gamma$ calculated in the MD simulation at high density, given the approximations involved in finding $\omega_{\rm D}$.

\begin{figure}
\begin{center}
{\scalebox{0.7}{\includegraphics{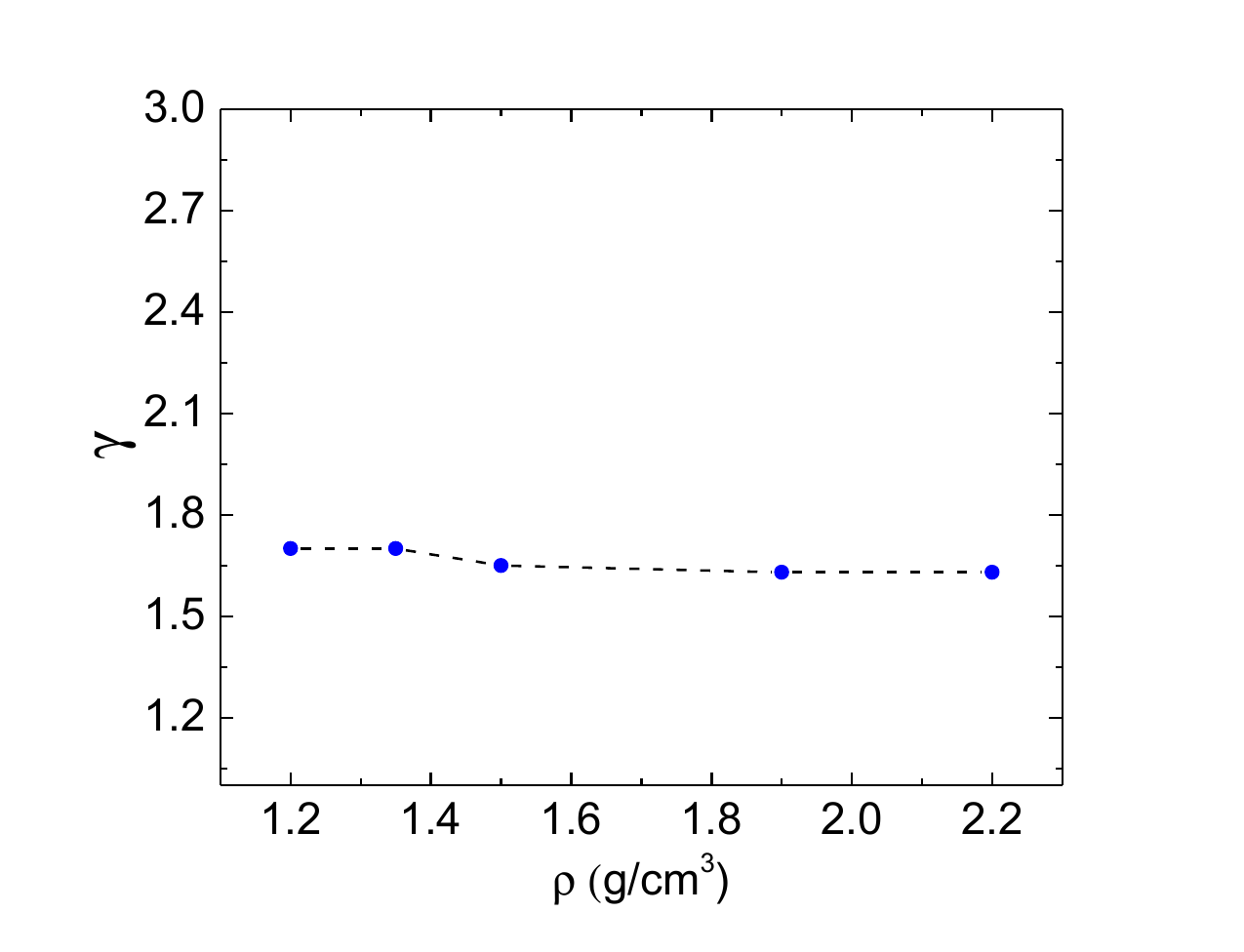}}}
\end{center}
\caption{Gr\"{u}neisen parameters at the Frenkel line for 5 different densities used in Figures \ref{low} and \ref{high}. $\gamma$ are plotted in the range approximately corresponding to the largest and smallest $\gamma$ in Figures \ref{low} and \ref{high}.}
\label{5points}
\end{figure}

We now address the behavior of $\gamma$ at the FL and plot the GP at all five densities and temperatures corresponding to the FL in Figure \ref{5points}. $\gamma$ are plotted in the range approximately corresponding to the largest and smallest $\gamma$ in Figures \ref{low} and \ref{high}. We observe that $\gamma$ is nearly constant at the FL: $\gamma=1.6-1.7$. This is an interesting result, given that the corresponding temperatures at the FL varies by more than an order of magnitude.

We propose the following explanation of the near constancy of $\gamma$ at the FL. The universality of $\gamma$ is related to scaling. At high energy (e.g. high pressure or temperature), particle interactions mostly involve the repulsive part of the potential. Therefore, the interatomic potential for Ar (as well as for many other systems) becomes effectively close to the soft-sphere potential $U\propto\frac{1}{r^n}$ \cite{stishov,hansen}, the classic example of a homogeneous potential. According to the Klein theorem \cite{klein1,ll,stishov-ufn}, the non-ideal part of the partition function depends on density $\rho$ and temperature as $\frac{\rho^{\frac{n}{3}}}{T}$ rather than on $\rho$ and $T$ separately. The resulting relationship between temperature and pressure at the melting line is $P_m\propto T_m^{1+\frac{3}{n}}$ \cite{stishov-ufn}. (Interestingly, the kinetic energy is also a homogeneous function of the second order, leading to scaling of kinetic coefficients such as viscosity and diffusion \cite{hiwatari,zhah}). Zhakhovsky extended the scaling argument \cite{zhah} and noted that, more generally, scaling always exists along those lines on the phase diagram where particle trajectories are similar or change in a similar way as they do at, for example, the melting line. Recall that the Frenkel line separates the combined oscillatory and diffusive motion below the line from purely diffusive motion above the line  \cite{ropp,pre,ufn,prl,phystoday}. Therefore, we expect the scaling relationship $P_{\rm F}\propto T_{\rm F}^{1+\frac{3}{n}}$ to hold at the FL as it does for the melting line. Such a relationship has been indeed ascertained in the soft-sphere system as well as LJ system at high pressure on the basis of MD simulations \cite{prl}. Then, $\gamma=V\frac{dP}{dE}=V\frac{dP}{dT}\frac{dT}{dE}\propto VT^{\frac{3}{n}}\frac{1}{C_V}$. Using the scaling relationship $V\propto T^{-\frac{3}{n}}$ from the Klein theorem, this gives $\gamma=f(n)\frac{1}{C_v}$, where $f(n)$ is the function of $n$ only. As mentioned earlier, $C_v$ is constant at the FL \cite{ropp,prl}. Hence, $\gamma$ at the FL does not depend on temperature and pressure, i.e. is a universal parameter for a system with a given $n$.

\begin{figure}
\begin{center}
{\scalebox{0.7}{\includegraphics{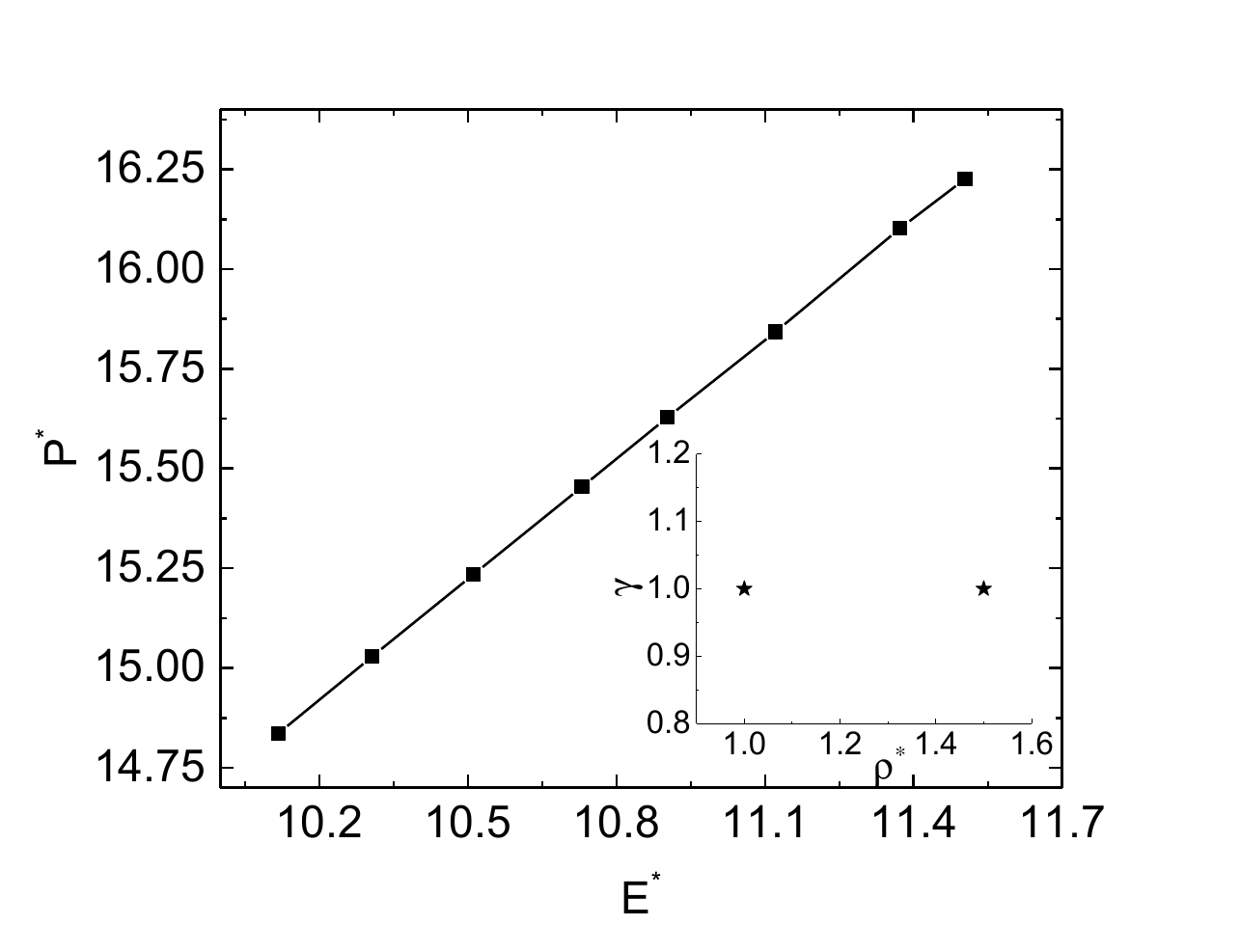}}}
\end{center}
\caption{The dependence of pressure on energy for the soft sphere system with $n=6$ at $\rho$*$=1.0$. Pressure and energy are shown in soft-sphere units. The inset shows the Gr\"{u}neisen parameter at two densities at the FL: $\rho$*$=1.0$ and $\rho$*$=1.5$.}
\label{softsphere1}
\end{figure}

To compare the results of the scaling argument with MD simulations further, we have calculated $\gamma$ for the soft-sphere system in a wide range of density and temperature. We show the results for $n=6$ in Figure \ref{softsphere1} for two different densities $\rho$*$=1.0$ and $\rho$*$=1.5$. Consistent with the scaling argument above (the soft-sphere system obeys the scaling argument) we observe that $\gamma$ is nearly constant at the FL.

We note that $\gamma$ for the soft-sphere system at the FL increases with $n$: using the previous data \cite{we-jept} we calculate $\gamma$ to be 1.5 for $n=12$ at the FL. This is close to $\gamma$ at the FL for the LJ system (see Figure \ref{5points}). This can be understood because the LJ potential becomes close to the soft-sphere potential at high pressure and high temperature as discussed above.

Before concluding, we make two remarks. First, we recall the earlier observation that $\gamma$ is constant along the isochore \cite{amoros}. This was related to a narrow range of pressure and temperature where the system can be approximated by a soft-sphere system with nearly constant effective radii and packing fraction and whose GP is constant along the isochore as mentioned earlier. At the same time, our results involving large range of pressure and temperature indicate that $\gamma$ can vary substantially, from those values typical of solids to the dense-gas ones.

Second, it will be interesting to evaluate the GP in the vicinity of the critical point. According to (3), $\gamma$ is governed by quantities which diverge at the critical point: compressibility, thermal expansion and heat capacity. Assuming, as is often done in the theory of critical phenomena, that the divergences of $\alpha$ and $\beta_T$ are equivalent, $\gamma$ at the critical point is governed by the behavior of $C_v$. For real systems, $C_v$ has a weak power divergence at the critical point, and $\gamma$ can be predicted to be close to 0. This point warrants further investigation.

\section{Conclusions}

In summary, we have calculated the Gr\"{u}neisen parameter of supercritical matter for two model systems in a very wide range of pressure and temperature. We find that $\gamma$ varies in a wide range which interestingly includes the solid-like values. We also find that $\gamma$ is nearly constant along the Frenkel line and rationalize this finding using the scaling of system properties along the lines where particle dynamic changes qualitatively. It is likely that a more general statement applies: any dimensionless parameter is universal at the line where scaling operates.

\section{Acknowledgments}

This research utilised Queen Mary's MidPlus computational facilities, supported by QMUL Research-IT and funded by EPSRC Grant EP/K000128/1. L.W., M.T.D., and K.T. are grateful to Royal Society and CSC. V.V.B. and Yu.D.F. are grateful to RSF (14-22-00093).


\end{document}